\documentclass[conference, table,dvipsnames]{IEEEtran}
\IEEEoverridecommandlockouts
\usepackage{cite}
\usepackage{amsmath,amssymb,amsfonts}
\usepackage{graphicx}
\usepackage{textcomp}

\usepackage{MnSymbol}

\usepackage{textcomp}
\usepackage{graphicx}
\usepackage{amssymb}
\usepackage{booktabs}
\usepackage{subcaption}
\usepackage{svg}
\usepackage{float}
\restylefloat{table}

\usepackage{booktabs}
\usepackage{graphicx}
\usepackage{bm}
\usepackage{hyperref}

\usepackage{multicol}
\usepackage{xcolor}
\usepackage{listings}

\usepackage{algorithm}
\usepackage{algpseudocode}
\usepackage{multirow,makecell}
\usepackage{units}

\usepackage{enumitem}
\usepackage{mathtools}

\usepackage{fancyhdr}

\DeclareMathOperator*{\argmin}{arg\,min}

\renewcommand{\vec}[1]{\boldsymbol{#1}}
\renewcommand{\th}{\boldsymbol{\theta}}
\newcommand{\x}{\vec{x}}
\newcommand{\y}{\boldsymbol{y}}
\newcommand{\Q}{\boldsymbol{Q}}

\newcommand{\zero}{\boldsymbol{0}}
\algrenewcommand\algorithmicrequire{\textbf{Input:}}
\algrenewcommand\algorithmicensure{\textbf{Output:}}

\definecolor{mylightblue}{HTML}{9FA8DA}
\definecolor{mygreen}{RGB}{144, 238, 144} % Light green
\definecolor{mypink}{RGB}{255, 182, 193} % Light pink/red color
\definecolor{customblue}{rgb}{0.9, 0.9, 1.0} % Define 
\definecolor{deepblueviolet}{RGB}{80,80,180}

\definecolor{lightsand}{RGB}{237,224,200}
\definecolor{sandtext}{RGB}{160,130,60}

\def\BibTeX{{\rm B\kern-.05em{\sc i\kern-.025em b}\kern-.08em
    T\kern-.1667em\lower.7ex\hbox{E}\kern-.125emX}}

\begin{document}

\title{Accelerated Spatio-Temporal Bayesian Modeling for Multivariate Gaussian Processes}

\author{
\IEEEauthorblockN{Lisa Gaedke-Merzhäuser\textsuperscript{*}}
\IEEEauthorblockA{\textit{KAUST} \\
Thuwal, Saudi Arabia \\
lisa.gaedkemerzhauser\\@kaust.edu.sa}
\and
\IEEEauthorblockN{Vincent Maillou\textsuperscript{*}}
\IEEEauthorblockA{\textit{ETH Zurich} \\
Zurich, Switzerland \\
vmaillou@iis.ee.ethz.ch}
\and
\IEEEauthorblockN{Fernando Rodriguez Avellaneda}
\IEEEauthorblockA{\textit{KAUST} \\
Thuwal, Saudi Arabia \\
fernando.rodriguezavellaneda\\@kaust.edu.sa}
\and
\IEEEauthorblockN{Olaf Schenk}
\IEEEauthorblockA{\textit{Università della Svizzera italiana} \\
Lugano, Switzerland \\
olaf.schenk@usi.ch}
\and
\IEEEauthorblockN{Mathieu Luisier}
\IEEEauthorblockA{\textit{ETH Zurich} \\
Zurich, Switzerland \\
mluisier@iis.ee.ethz.ch}
\and
\IEEEauthorblockN{Paula Moraga}
\IEEEauthorblockA{\textit{KAUST} \\
Thuwal, Saudi Arabia \\
paula.moraga@kaust.edu.sa}
\and
\IEEEauthorblockN{Alexandros Nikolaos Ziogas}
\IEEEauthorblockA{\textit{ETH Zurich} \\
Zurich, Switzerland \\
alziogas@iis.ee.ethz.ch}
\and
\IEEEauthorblockN{Håvard Rue}
\IEEEauthorblockA{\textit{KAUST} \\
Thuwal, Saudi Arabia \\
haavard.rue@kaust.edu.sa}
}

% \author{
%   \IEEEauthorblockN{Lisa Gaedke-Merzhäuser\IEEEauthorrefmark{1},
%                     Vincent Maillou\IEEEauthorrefmark{2},
%                     Fernando Rodriguez Avellaneda\IEEEauthorrefmark{1},
%                     Olaf Schenk\IEEEauthorrefmark{3},
%                     Mathieu Luisier\IEEEauthorrefmark{2}}
%   \\
%   \IEEEauthorblockA{\IEEEauthorrefmark{1}Institution One \\
%                     Email: author.one@example.com, author.three@example.com}
%   \\
%   \IEEEauthorblockA{\IEEEauthorrefmark{2}Institution Two \\
%                     Email: author.two@example.com}
%       \\
%   \IEEEauthorblockA{\IEEEauthorrefmark{3}Institution Two \\
%                     Email: author.two@example.com}
% }

\maketitle

% \begingroup
% \renewcommand\thefootnote{}
% \footnotetext{\textsuperscript{*}These authors contributed equally to this work.}
% \endgroup

\begin{abstract}
Multivariate Gaussian processes (GPs) offer a powerful probabilistic framework to represent complex interdependent phenomena. 
They pose, however, significant computational challenges in high-dimensional settings, which frequently arise in spatial-temporal applications.
We present DALIA, a highly scalable framework for performing Bayesian inference tasks on spatio-temporal multivariate GPs, based on the methodology of integrated nested Laplace approximations. Our approach relies on a sparse inverse covariance matrix formulation of the GP, puts forward a GPU-accelerated block-dense approach, and introduces a hierarchical, triple-layer, distributed memory parallel scheme.
We showcase weak scaling performance surpassing the state-of-the-art by two orders of magnitude on a model whose parameter space is 8$\times$ larger and measure strong scaling speedups of three orders of magnitude when running on 496 GH200 superchips on the Alps supercomputer.
Applying DALIA to air pollution data from northern Italy over 48 days, we showcase refined spatial resolutions over the aggregated pollutant measurements.
\end{abstract}


\begin{table*}[t]
\centering
\resizebox{\textwidth}{!}{
\begin{tabular}{@{}lccccccccc@{}}
\toprule
 \textbf{Framework} & \textbf{Modeling} & \multicolumn{3}{c}{\textbf{Parallelism}} & \multicolumn{4}{c}{\textbf{Implementation}} \\
\cmidrule(lr){3-5} \cmidrule(lr){6-9}
 & & $f_{\text{obj}}$ & $Q_p/Q_c$ & Solve & GPU & Comm & Scaling & Language \\
\midrule
R-INLA & Extensive & SM & - & PARDISO (SM) & - & - & Single-node & C, R \\
$\text{INLA}_{\text{DIST}}$ & ST & \colorbox{mypink!60}{DM} & \colorbox{mypink!60}{DM} & BTA Solver (SM) & Solver-only & \colorbox{mypink!60}{MPI} & 18 GPUs & C++, CUDA \\
DALIA & ST\colorbox{mygreen!60}{+COREG} & \colorbox{mypink!60}{DM} & \colorbox{mypink!60}{DM} & \colorbox{mygreen!60}{BTA Solver (DM)} & \colorbox{mygreen!60}{Entire framework} & \colorbox{mypink!60}{MPI}\colorbox{mygreen!60}{+NCCL} & \colorbox{mygreen!60}{496 GPUs} & Python, CuPy \\
\bottomrule
\end{tabular}
}
\footnotesize\\
\vspace{0.5em}
\textbf{Notes}: SM = Shared Memory; DM = Distributed Memory; ST = Spatio-Temporal; COREG = Coregional.

\caption{Comparison of relevant INLA implementations, highlighting modeling, parallelism, and implementation features. Innovations in \textcolor{mygreen}{green} are introduced by our framework; those in \textcolor{mypink}{red} are improvements or shared features with INLA\textsubscript{DIST}.}
\label{tab:overview_implementation}
\end{table*}

\section{Introduction}
\label{sec:intro}

\begin{figure}[t]
    \centering
    \includegraphics[width=\columnwidth]{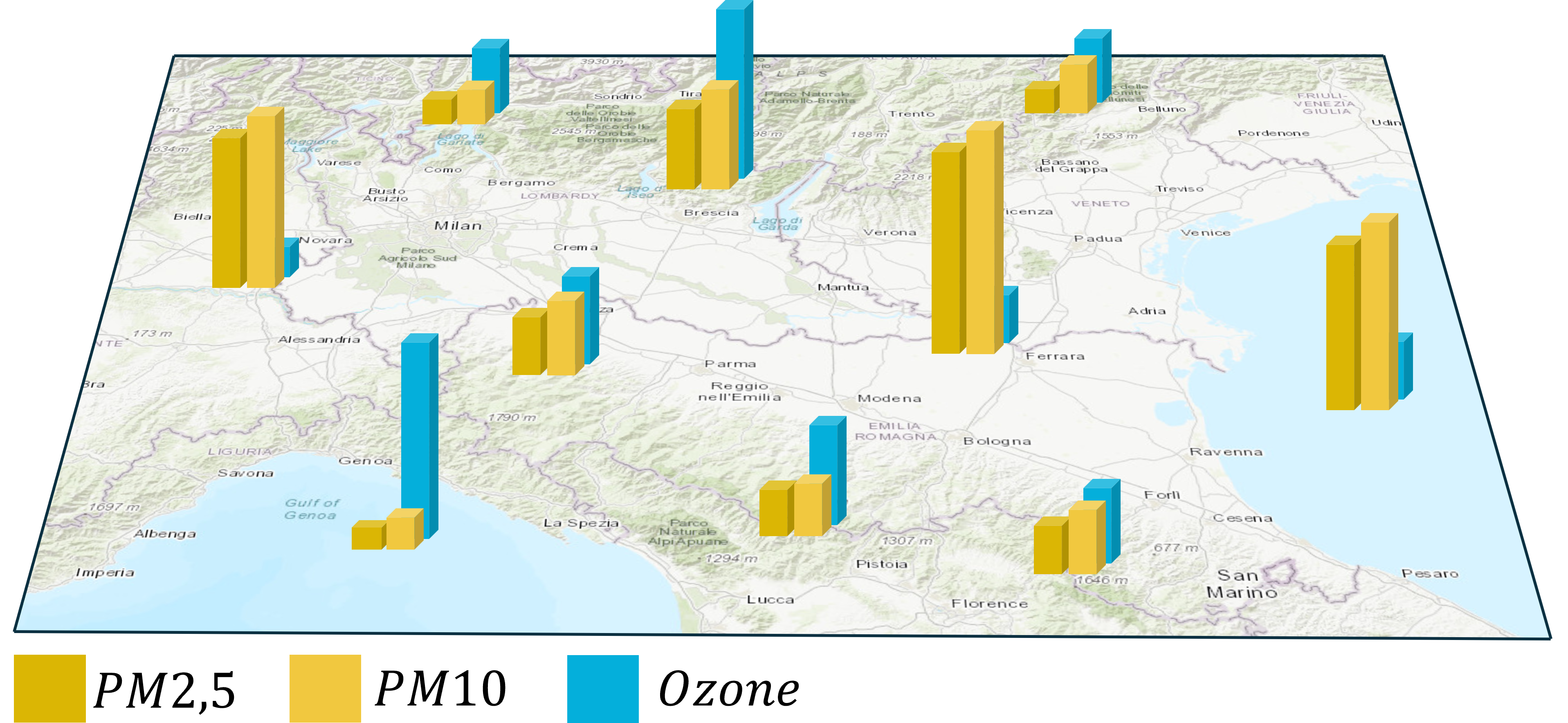}
    \caption{Representation of pollutant measurements (PM$_{2.5}$, PM$_{10}$, and $O_3$) 
    over northern Italy.}
    \label{fig:intro_polution}
\end{figure}

Air pollution poses a significant public health risk \cite{kelly2015air}. 
The exposure to pollutants such as ozone ($O_3$) and particulate matter (e.g., PM$_{2.5}$, PM$_{10}$) has been linked to increased mortality and a wide range of health conditions, including respiratory infections, cardiovascular diseases, and lung cancer~\cite{cohen2017, pope2011lung}. 
Accurately modeling the spatio-temporal distribution of air pollutants is essential to improving exposure assessments.
Being able to quantify model uncertainty in this context also plays an important role as interest extends beyond the average pollutant concentrations at fixed locations and time points.
Instead, information on the range of likely values over continuous time periods and uncertainties in the measurements can be central to determining risks of continuously exceeding regulatory thresholds. Subsequently, this can support better informed policy making and public health protection.
Fig.~\ref{fig:intro_polution} visualizes selected measurements of air pollutants over northern Italy.

\begingroup
\renewcommand\thefootnote{}
\footnotetext{\hfill\textsuperscript{*}These authors contributed equally to this work.}
\endgroup

Bayesian statistics offers a framework where all unknown parameters are assumed to be random variables with associated probability distributions, instead of fixed unknowns. Through the application of Bayes' rule, this allows for the seamless incorporation of prior knowledge and a natural way to quantify uncertainty.
A powerful class of priors used to represent the latent or unknown process of interest is Gaussian processes (GPs) \cite{rasmussen2006gaussian}, whose parameters and hyperparameters are estimated throughout the inference procedure. 
A comprehensive Bayesian approach comes, however, at an increased computational cost. 
Scalability becomes a major challenge for all Bayesian methods \cite{blei2017variational} when the number of model parameters, available data, and model complexity increases.
Sampling-based approaches like Markov Chain Monte Carlo (MCMC) methods suffer from slow convergence for high-dimensional parameter spaces due to slow mixing of the Markov chain, requiring large numbers of iterations \cite{martino2019inla}.
Approximation-based approaches like variational inference rely on simplified assumptions to achieve better scalability, which can, unfortunately, come at the expense of posterior approximation accuracy \cite{blei2017variational}.
Integrated nested Laplace approximations (INLA), another approximation-based inference method, offer the potential for better scalability but are more restrictive in their applicable model class \cite{rue2009approximate}.
INLA is, however, applicable to multivariate GPs, making it a suitable choice in this context for providing fast yet accurate inference results compared to other Bayesian approaches~\cite{opitz2017latent}.
Its reference implementation is available via the R-INLA package \cite{inlapkg}, which is particularly popular among applied statisticians due to its user-friendly R interface.
It has been employed in a variety of studies on air pollution and other environmental phenomena, see e.g. \cite{forlani2020joint, zhong2023spatial, rodriguez2025multivariatedisaggregationmodelingair, gomez2019multivariate}.
An overview of its computational features from an implementation perspective is provided in Table~\ref{tab:overview_implementation}.
%Nonetheless, scalability remains a bottleneck as these methods scale with the number of observations or parameters.
Since the computational cost of the INLA methodology is inherently tied to the number of unknown model parameters, many large-scale models continue to pose computational challenges.
% Since the computational cost of the INLA methodology is inherently tied to the number of unknown model parameters, and therefore many large-scale models continue to pose computational challenges.
Among these, a particularly challenging class is spatial and spatio-temporal models, whose representation over the spatial or spatio-temporal domain requires many parameters and is often accompanied by large numbers of observations.
The INLA\textsubscript{DIST} library \cite{gaedkeIntegrated2024, inladist_git} addresses these increased computational demands by offering a GPU-accelerated implementation specifically tailored to the characteristics of univariate spatial and spatio-temporal models, see Table~\ref{tab:overview_implementation} for more details. 

For many applications, it is, however, not sufficient to only consider a univariate process with a single response variable; instead, multiple correlated processes are required. 
Air pollution levels are, for example, often measured as the joint combination of particulate matter, such as PM$_{2.5}$ and PM$_{10}$, and ozone concentration, which are interdependent quantities.

A common approach to modeling multivariate spatial processes is the linear model of coregionalization (LMC), which represents the response variable as a linear combination of independent univariate latent Gaussian processes~\cite{Grzebyk2007MultivariateAA}. Initially developed for spatial settings, the LMC approach extends naturally to spatio-temporal applications, particularly in environmental data analysis~\cite{DeIaco2003}. 
%The LMC offers flexibility, as its covariance structure can be adapted to different problems, and interpretability, as it captures linear dependencies between response variables through model parameters. 
From a practical perspective, however, many studies limit themselves to spatial analyses \cite{moraga2020species,pavani2023joint, ferreira2006spatial}, as the computational cost and corresponding runtimes using available software libraries for spatio-temporal models become prohibitively high \cite{Cappello2022, alie2024computational}.
This increased computational cost is induced by the high-dimensional latent parameter space and the operations required on its associated covariance or inverse covariance (precision) matrices. 
As the representation of spatio-temporal processes already leads to high-dimensional parameter spaces, combining multiple such univariate processes, only further exacerbates the computational complexity.

To overcome these limitations, we propose a novel framework, DALIA, to perform large-scale spatio-temporal Bayesian modeling for multivariate Gaussian processes. 
Our approach builds on the sparse Gaussian Markov random field representation of the underlying latent Gaussian processes' precision (inverse covariance) matrices. 
We use a spatio-temporal stochastic partial differential equation (SPDE) approach to represent the univariate GPs using a computationally efficient finite element discretization.
The latter is combined with an efficient multivariate LMC formulation that maintains sparsity in the joint precision matrices.
In turn, this enables us to employ specialized structured sparse linear solvers to handle the computational bottleneck operations: Cholesky decomposition, triangular solve, and selected inversion.

The main contributions of this work are the following:
\begin{itemize}
    \item Computationally advantageous expression of multivariate coregional models;
    \item Implementation of a hierarchical, nested, three-layer distributed memory parallelization scheme;
    \item A time-domain decomposition of the spatio-temporal models allowing the integration of a distributed linear solver;
    \item The development of a GPU-accelerated distributed-memory triangular solve routine;
    \item Performance evaluations of our framework and scaling analysis on up to 496 GH200 superchips on the Alps supercomputer in the Swiss National Supercomputing Center (CSCS), achieving a three orders of magnitude speedup over the state-of-the-art;
    \item Application of our framework to a large-scale multivariate spatio-temporal GP model on air pollution providing refined spatial resolutions over aggregated measurements;
\end{itemize}

Our approach for multivariate spatio-temporal Bayesian models goes beyond the existing state of the art, in both achievable problem size and time-to-solution.
The remainder of this paper is organized as follows. In Section~\ref{sec:background}, we introduce the necessary statistical background for the formulation of multivariate, spatio-temporal Gaussian processes.
%The efficient inference and computation of these models are the main focus of this work.
In Section~\ref{sec:inla}, we introduce the INLA methodology and discuss its current implementations. Our framework and its novelties are presented in Section~\ref{sec:DALIAframework}, and numerical evaluations are performed in Section~\ref{sec:results}.
We present the application of our framework to an air pollution application over northern Italy in Section~\ref{sec:scientific_results}.
Finally, conclusions are drawn in Section~\ref{sec:conclusion}.

\section{Background}
\label{sec:background}

% introduce table
Here, we introduce the statistical model formulation, the inference method, and the associated computational operations. Table~\ref{tab:notation} summarizes the notation and abbreviations used throughout this document.

\subsection{Univariate Gaussian Processes}
%\subsubsection{Gaussian Process Priors}

A GP prior generalizes the idea of placing a multivariate normal prior over a finite-dimensional space to the infinite-dimensional setting, enabling the modeling of a continuous process through evaluations at a finite set of points. 
In spatial and spatio-temporal settings, we can leverage the connection between GPs and sparse Gaussian Markov Random fields \cite{rue2005gaussian}, which allows us to define a multivariate normal prior with a sparse precision (inverse covariance) matrix to represent the GP. 
The corresponding probability density function is given by
\begin{equation}
p(\mathbf{x} | \th) = (2\pi)^{-\frac{n}{2}} |\Q_p(\th)|^{\frac{1}{2}} \exp\left(-\frac{1}{2} \mathbf{x}^T \Q_p(\th) \mathbf{x}\right).
\label{eq:GMRF}
\end{equation}
For $\x = [x_1, x_2, \dots, x_n]$, each variable $x_j$ is associated with a spatial or spatio-temporal location $s_j$, such that $x_j = x(s_j)$.
The sparse symmetric positive definite precision matrix $\Q_p(\th)$ is parametrized by the hyperparameters $\th$ that govern the spatial or spatio-temporal dependencies. For brevity, we will implicitly assume this dependency on $\th$ from now on and omit in the notation. 
%The determinant of $\Q_p$ is denoted by $|\Q_p|$.
Being able to evaluate Eq.~\ref{eq:GMRF} efficiently for fixed values of $\th$ and $\x$ is essential to the overall inference process and arises repeatedly for different parameter configurations.
It requires the factorization of $\Q_p$ to compute its determinant, denoted by $\vert \cdot \vert$, from the diagonal entries of the Cholesky factor.
The size of $\x$ and $\Q_p$ grows with increasing numbers of time steps or refined spatial resolutions.

\begin{table}[t]
\scriptsize
\centering
\rowcolors{2}{gray!15}{white} % Alternating row colors start from the second row
\begin{tabular}{p{1.78cm}p{6.25cm}}
\toprule
\textbf{Name} & \textbf{Description} \\
\midrule
% First section: Matrix Parameters
\rowcolor{white} \multicolumn{2}{c}{\textbf{Statistics-related Terms}} \\ 
\cmidrule(lr){1-2} % Midrule spanning only the description column
INLA & Integrated Nested Laplace Approximation methodology. \\
$Q_p$ & Prior precision matrix, exhibits a block-tridiagonal sparsity pattern.\\
$Q_c$ & Conditional precision matrix, exhibits a block-tridiagonal with arrowhead sparsity pattern.\\
% Second section: Algorithm Concepts
\cmidrule(lr){1-2}
\rowcolor{white} \multicolumn{2}{c}{\textbf{Spatio-Temporal Models Parameters}} \\ 
\cmidrule(lr){1-2} % Midrule spanning only the description column
$n_t$ & Number of time-steps in the time-domain.\\
$n_s$ & Number of mesh nodes in the spatial-domain.\\
$n_r$ & Number of fixed effects.\\
$n_v$ & Number of response variables (joint ST models).\\
% Third section: Matrix Types and Algorithms
\bottomrule
\end{tabular}
\caption{Symbols and terms used in this work.}
\label{tab:notation}
\end{table}

\subsubsection{Sparse precision matrices, spatio-temporal modeling and the SPDE Approach}

Precision matrices encode dependencies between neighboring variables. They are often sparse as $Q_{ij} = 0$ if and only if $x_i$ and $x_j$ are conditionally independent given all other variables. This stands in contrast to covariance (inverse precision) matrices, which capture overall dependency structures and are therefore generally dense. 
%The sparse precision matrix representation bears computational and memory advantages.
To construct $\Q_p$ we rely on the stochastic partial differential equation (SPDE) approach 
\cite{lindgren2011explicit, lindgren2022diffusion}, which expresses the spatial and spatio-temporal dependencies between variables as the solution to a SPDE. 
This reformulation can be efficiently discretized using finite elements, preserves desirable statistical model properties, easily accommodates unstructured observation locations and complicated spatial domains while simultaneously preserving sparsity~\cite{lindgren2022spde}.
The hyperparameters $\th$ regulate the characteristics of the SPDE, such as spatial or temporal correlation range, which define the maximum distance over which significant spatial or temporal correlation is present, respectively.
For spatio-temporal models, ordering the variables sequentially over time, one obtains a structured block-tridiagonal (BT) sparsity pattern in $\Q_p$, where each diagonal block relates to the spatial discretization of a single time step, where the time steps are coupled through the off-diagonal blocks (see Fig.~\ref{fig:permutation_coreg}a) for illustration). 
This allows for capturing dynamics in the Gaussian process that are specific to distinct regions in space and time, such as varying pollutant concentrations influenced by increased industrial activities or particular geographical features. These types of region- or time-specific variations are also referred to as random effects.

\subsubsection{Bayesian model}

In addition, we consider fixed effects in our model formulation to account for factors that have a consistent influence on the process, such as elevation or distance to coastline. 
We incorporate them in the formulation of Eq.~\ref{eq:GMRF}, by extending $\x$ and $\Q_p$. 
Together, they describe the following linear model for the observations
\begin{equation}
\vec{y} = \vec{A} \x + \vec{\epsilon}, 
%=  \vec{A}_{st} \vec{x}_{st} + \vec{A}_f \vec{\x}_f + \vec{\epsilon}, 
\label{eq:univariate_process}
\end{equation}
where $\y = [y_1, \dots, y_m]$ is the observed data, i.e. $y_i$ is the concentration of a particular pollutant at a specific space-time location, $\x = [\x_{st},\x_f]$ are the spatio-temporal random effects, and the fixed effects, respectively.
$\vec{A}$ is a sparse design matrix linking the effects to the response and $\vec{\epsilon}$ represents an error term to account e.g., for measurement noise. 

\subsubsection{Conditional Gaussian distribution}
Building on the underlying GP prior, the second high-dimensional multivariate normal distribution of interest is a conditional distribution formed by combining the GP prior with a second-order Taylor expansion of the likelihood that describes the distribution of the data.
We denote it by 
\begin{equation}
    p_G(\x) = (2\pi)^{-\frac{n}{2}} |\Q_c|^{\frac{1}{2}} \exp\left(-\frac{1}{2} (\mathbf{x} - \vec{\mu})^T \Q_c (\mathbf{x} - \vec{\mu}\right).
    \label{eq:p_G}
\end{equation}
The approximation $p_G$ is exact if the likelihood is itself Gaussian.
Here, $\vec{\mu}$ denotes its conditional mean.
Its precision matrix is constructed as
\begin{equation}
\Q_c = \Q_c(\th) =  \Q_p(\th) + \vec{A}^T \vec{D}(\th) \,\vec{A}
\label{eq:Q_cond}
\end{equation}
where $\vec{D}$ is the negative Hessian of the log-likelihood for fixed quantities $\th$ and $\x$. The matrix $\vec{A}$ is the design matrix from Eq.~\ref{eq:univariate_process}. 
The sparsity pattern of $\Q_c$ is closely related to $\Q_p$, for the spatio-temporal part $\vec{A}$ preserves its sparsity pattern. For the part related to the fixed effects, additional non-zeros are introduced, resulting in a block-tridiagonal arrowhead (BTA) matrix.
While $\Q_c$ stores information on the conditional dependencies, its inverse $\vec{\Sigma}_c = \Q^{-1}_c $ contains information on the linear dependence between two variables in the off-diagonal entries, while its diagonal entries store the variances of the individual variables which are of interest for providing uncertainty estimates.
From a computational perspective this requires the computation of selected entries of the full inverse of $\Q_c$, more specifically those entries that are nonzero in $\Q_c$. 

% The previous section describes a univariate process. 

\subsection{Multivariate Gaussian Processes}
\label{subsec:MGP}
%\subsubsection{Linear Models of Coregionalization}

To represent multiple dependent spatio-temporal processes, we rely on a linear model of coregionalization (LMC)\cite{journel1978mining, schmidt2003bayesian} approach. 
LMC are widely used to represent multivariate spatial or spatio-temporal processes \cite{cappello2021computational, pellegrino2020spatio} and construct a joint multivariate process by combining $n_v$ independent univariate latent processes through a coregionalization matrix. While this formulation holds for any number of response variables $n_v$, for ease of notation we consider the case $n_v = 3$. The multivariate linear model then becomes

{\footnotesize
\begin{equation}  
{%\footnotesize %\small
\underbrace{
\begin{bmatrix}
\vec{y}_1 \\
\vec{y}_2 \\
\vec{y}_3
\end{bmatrix}
}_{\vec{y}}
=
\underbrace{
\begin{bmatrix}
\sigma_1 \vec{I} & \vec{0} & \vec{0} \\
\lambda_1 \sigma_1 \vec{I} & \sigma_2 \vec{I} & \vec{0} \\
(\lambda_3 + \lambda_1 \lambda_2) \sigma_1 \vec{I} & \lambda_2 \sigma_2 \vec{I} & \sigma_3 \vec{I}
\end{bmatrix}
}_{\boldsymbol{\Lambda}}
\underbrace{
\begin{bmatrix}
\vec{A}_1 & 0 & 0 \\
0 & \vec{A}_2 & 0 \\
0 & 0 & \vec{A}_3
\end{bmatrix}
}_{\vec{A}}
\cdot
\underbrace{
\begin{bmatrix}
\vec{x}_1 \\
\vec{x}_2 \\
\vec{x}_3
\end{bmatrix}
}_{\vec{x}}
+
\vec{\epsilon}
}. % end small
\label{eq:multivariate_model}
\end{equation}
}
Here each $\y_i$ contains the observations related to a different response variable. 
The matrix $\vec{\Lambda}$ consists of the scale parameters $\sigma_i$ and the coupling terms $\lambda_i$ which encode the interactions between the univariate processes $\x_1, \x_2, \x_3$.
The resulting covariance matrix of the multivariate process is given by
\begin{equation}
\vec{\Sigma}_{n_v} =  \vec{\Lambda} \
\left[
\begin{array}{ccc}
  \vec{\Sigma}_{1} & \zero & \zero \\
  \zero &  \vec{\Sigma}_{2} &  \zero  \\
    \zero &   \zero &  \vec{\Sigma}_{3}
\end{array}
\right] \
\vec{\Lambda}^T,
\label{eq:covariance_LMC}
\end{equation}
with the block diagonal matrix containing the spatio-temporal covariance matrices $\vec{\Sigma}^i_p = (\Q^i_p)^{-1}$ of the univariate processes $\x_i$ which are assumed to be conditionally independent and have unit variance. 
From this we derive the corresponding sparse precision matrix formulation and a permutation scheme to recover a BTA sparsity pattern in Sec.~\ref{sec:DALIAframework}.

\begin{figure*}[t]
    \centering
    \includegraphics[width=\textwidth]{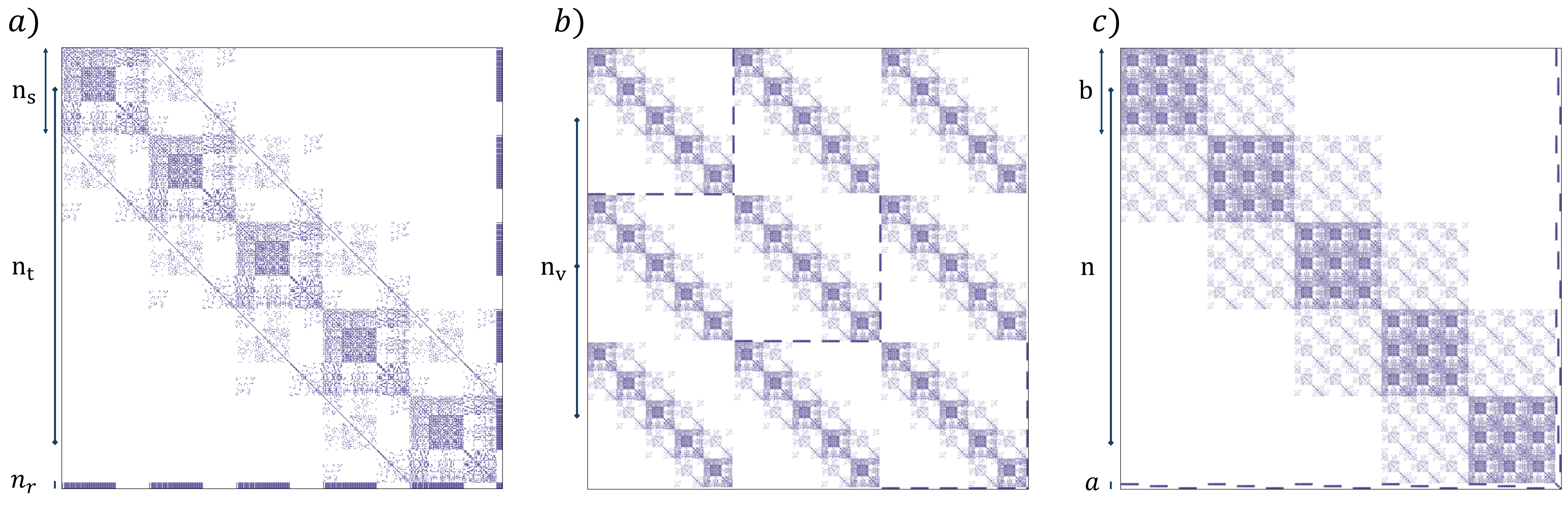}
    \caption{Conditional precision matrices of $a)$ a univariate spatio-temporal model, $b)$ a trivariate spatio-temporal, coregional, model and $c)$ its permuted block-tridiagonal with arrowhead version. The permuted matrix c) is a block-tridiagonal with arrowhead structured sparse matrix, with $n$ main diagonal blocks, their size $b$, and the arrow tip block size $a$.}
    \label{fig:permutation_coreg}
\end{figure*}

\section{The INLA Methodology}\label{sec:inla}

The methodology of integrated nested Laplace approximations provides a way to approximate the posterior distributions of the unknown model parameters by combining the GP priors and the priors of the hyperparameters with the observed data using Bayes' rule~\cite{rue2009approximate}. 
From the above, other quantities of interest, like predictive distributions at unobserved locations or measures of uncertainty, can be derived.
%Integrated nested Laplace approximations (INLA) is a methodology for performing fast and accurate Bayesian inference tasks, applicable to the class of latent Gaussian models, a subclass of additive Bayesian hierarchical models comprising Gaussian processes, regression, auto-regressive, and spatial and spatio-temporal models.
The INLA methodology is routinely employed across a wide variety of domains, including epidemiology \cite{rodriguez2024estimating}, environmental sciences \cite{huang2017enviromental}, ecology \cite{moraga2020species}, finance \cite{dutta2024modeling}, and social sciences \cite{coleman2024exploring}.
% potentially put this in intro
The main idea behind INLA is to form a (nested) approximation of the posterior distributions of interest, centered around the mode of the hyperparameters, which first needs to be identified through numerical optimization. 
%From a computational perspective INLA is 

\subsubsection{Approximation of the joint posterior}

A fundamental building block is the construction of the joint posterior of the hyperparameters given the observations $\y$. Derived using Bayes' rule it is formulated as 
\begin{align}
p(\th | \y) \propto \frac{p(\th) p(\x | \th) \ell(\y | \th, \x)}{p_G(\x | \th, \y)} %\bigg\rvert_{\x = \x*(\th)}, 
\label{eq:post_theta}
\end{align}
where $p(\th)$ denotes the prior of the hyperparameters, $p(\x | \th)$ the GMRF approximation of the GP prior, $\ell(\y | \th, \x)$ the likelihood of the data. The denominator consists of a Gaussian approximation to the true conditional distribution $p(\x | \th, \y)$.
Taking the logarithm, we define the following objective function
% \begin{align}
%     f_{\text{obj}}(\th) := \text{log} \,  p(\th) + \text{log} \,  \ell(\y | \th, \x) + \text{log} \,  p(\x | \th) - \text{log} \, p_G(\x | \th, \y).
%     \label{eq:f_theta}
% \end{align}
\begin{equation}
\begin{aligned}
%{\small
    % f_{\text{obj}}(\th) \!:= \!\colorbox{lightsand}{$\!\text{log} \, p(\th)\!$}\!+ \!\colorbox{lightsand}{$\!\text{log} \, \ell(\y | \th, \x)\!$}\! 
    % + \!\colorbox{customblue}{$\!\text{log} \, p(\x | \th)\!$}\! - \!\colorbox{customblue}{$\text{log} \, p_G(\x | \th, \y)\!$}.
%}
f_{\text{obj}}(\th) :&= \colorbox{lightsand}{$\text{log} \, p(\th)$} + \colorbox{lightsand}{$\text{log} \, \ell(\y | \th, \x)$} \\
&+ \colorbox{customblue}{$\text{log} \, p(\x | \th)$} - \colorbox{customblue}{$\text{log} \, p_G(\x | \th, \y)$}.
\end{aligned}
\label{eq:f_theta}
\end{equation}
All subcomponents of $f_{\text{obj}}(\th)$ can be evaluated point-wise, while its closed-form expression is generally not available. 
While the first two terms are generally computationally \textcolor{sandtext}{cheap} to evaluate, the last two terms, are, \textcolor{deepblueviolet}{expensive} to compute. They represent the multivariate normal distributions presented in Eq.~\ref{eq:GMRF} and Eq.~\ref{eq:p_G}, and whose evaluations necessitate high-dimensional precision matrix factorizations and triangular solves for the latter.
We also note that since $p_G$ is exact for Gaussian likelihoods, $\Q_c$, the evaluation of the last two terms is parallelizable, i.e., $\Q_p$ and $\Q_c$ can be factorized in parallel.

\subsubsection{Optimization over the hyperparameters.}
%TODO: comment on uni-modal ...
Using Eq.~\ref{eq:f_theta}, INLA estimates the optimum of $f_{\text{obj}}(\th)$, and thus, the mode of $p(\th | \y)$ by solving the unconstrained optimization problem 
\begin{equation}
   \th^* =  \argmin_{\th} \ - f_{\text{obj}}(\th).
   \label{eq:optimization_problem}
\end{equation} 
using a BFGS algorithm~\cite{nocedal2006numerical}. In each iteration $l$, the quasi-Newton method requires the current function value $f_{\text{obj}}(\th^l)$ and its gradient $\nabla f_{\text{obj}}(\th^l)$ to propose the next step. As the analytical gradient is not easily available,
%, and automatic differentiation would require the differentiation through the large Cholesky factorizations, which is extremely inefficient from a computational perspective
a finite difference scheme is used to approximate the gradient, i.e., for each directional derivative at iteration $\th^l$, one computes
\begin{equation}
\frac{\partial}{\partial \theta_i} f_{\text{obj}}(\th^l) \approx \frac{f_{\text{obj}}(\th^l + h \vec{e}_i) - f_{\text{obj}}(\th^l - h \vec{e}_i)}{2h} \quad \text{for all } i.
\label{eq:FD_scheme}
\end{equation}
Here, $h$ is a small scalar, while $\vec{e}_i$ is the unit vector selecting the $i$-th component of $\th$.
This finite difference approximation bears the computational advantage that all gradient function evaluations $f_{\text{obj}}(\th^l + h \vec{e}_1), f_{\text{obj}}(\th^l - h \vec{e}_1), f_{\text{obj}}(\th^l + h \vec{e}_2), \dots$, can be performed in parallel within the same BFGS iteration.
Performing the optimization to identify $\th^*$ constitutes the computationally most expensive part of the method as it requires many repeated function evaluations of $f_{\text{obj}}$. 

\subsubsection{Marginal posterior distributions of the hyperparameters}
To quantify the uncertainty in the estimation of the hyperparameters, the negative Hessian of $f$ at the mode $\th^*$ is computed using a second-order finite difference scheme, similar to Eq.~\ref{eq:FD_scheme}. This allows for a Gaussian approximation of their marginal distributions. Alternatively, an eigenvector decomposition of the Hessian is used for reparametrization to accommodate skewness in the posterior marginals. This strategy is more accurate but computationally more expensive, requiring further evaluations of $f_{\text{obj}}(\th)$~\cite{martins2013bayesian}.

\subsubsection{Marginal posterior distributions of the latent parameters}

The posterior distributions of the latent Gaussian process are obtained using the Laplace approximations $p_G(\x | \th, \y)$, at different $\th$. In the simplest scenario, $p_G$ is only computed at the mode of $\th^*$. The marginal means $\mu_j$ are then directly deduced from $p_G(\x | \th^*, \y)$. 
% The marginal variances are obtained from calculating selected entries of the inverse of $\Q_c(\th^*, \x^*)$, necessitating its selected inversion.
The marginal variances are obtained from calculating selected entries of the inverse of $\Q_c$. This last operation required a specific set of challenging numerical methods, called \textit{selected inversion}. 

% These methods allow to recover selected entries of the inverse of a sparse matrix while avoiding the introduction of dense transcients results.

\subsection{Computational Overview}
\label{sec:scaling_challenges}

% From a computational perspective we identify three key steps within the INLA methodology
From a computational standpoint, the INLA methodology relies on 

\begin{enumerate}
\item\label{item:item1} Solving the optimization problem posed in Eq.~\ref{eq:optimization_problem} using an iterative quasi-Newton method. During each iteration $l$, the gradient can be computed in parallel using a finite difference scheme with parallel function evaluations of $f_{\text{obj}}$. Within each evaluation of $f_{\text{obj}}$ the factorization of the sparse matrices $\Q_p(\th^l), \Q_c(\th^l, \x^l)$ and a triangular solve are necessary. For Gaussian data, the two matrices are independent of each other and can be factorized in parallel.
\item\label{item:item2} The Hessian computation of $f_{\text{obj}}$ at the mode $\th^*$ using a second-order finite difference scheme, necessitating further parallelizable function evaluations of $f_{\text{obj}}$. 
% Thus, similar to~\ref{item:item1}.
\item\label{item:item3} Quantification of the uncertainty in the latent Gaussian process through the selected inversion of $\Q_c$ in different values of $\th$.
\end{enumerate}

% This implies that the scalability of the overall method strongly relies on leveraging parallelism wherever possible and efficiently handling the computational core operations -- Cholesky decomposition, triangular solve and matrix selected inversion.

This implies that the scalability of the overall method strongly relies on leveraging the possible opportunities for parallelism as well as efficiently handling the computational core operations; Cholesky decomposition, triangular solve and matrix selected inversion.

\subsubsection{Cholesky factorization and triangular solve}
Step~\ref{item:item1} and~\ref{item:item2} necessitate the recurrent computation of the Cholesky factors of $\Q_p$ and $\Q_c$.
The values of the non-zero entries in these precision matrices depend on the value at the current iteration $l$ of the hyperparameters $\th^l$. However, their sparsity structures (respectively BT and BTA) remain the same.
This allows us to employ taylored structured sparse solvers that leverage these characteristics.

\subsubsection{Selected inversion}

% These methods allow to recover selected entries of the inverse of a sparse matrix while avoiding the introduction of dense transcients results.

% While the inverse of a sparse matrix is dense, s
Selected inversion is a numerical method for computing selected entries of the inverse of a sparse matrix, while avoiding the introduction of dense transient results. 
As the size of the precision matrices arising in spatio-temporal models can reach several millions, dense transients would quickly make the operation unfeasible from both a computational and a memory standpoint.
From a method standpoint, only the entries that match the sparsity pattern of the precision matrix are needed, enabling the usage of selected inversion.
For general sparse matrices, an approach based on Cholesky factorization was first proposed in~\cite{takahashi_sellinv} with an implementation found in ~\cite{pardiso_1}. 
In the case of BT and BTA matrices, several methods have been developed~\cite{Kuzmin13, gaedkeIntegrated2024} and recent work has extended these algorithms to distributed-memory parallelism~\cite{maillou2025serinvscalablelibraryselected}.

\subsection{State-of-the-art Implementations of the INLA Methodology}

The reference implementation of the INLA methodology is through the R-INLA package \cite{inlapkg} and supports a wide range of different models including multivariate GP models and allows for an LMC approach.
The package is primarily written in C but provides an R interface to facilitate usage for applied statisticians. 
It leverages various strategies for performance enhancement~\cite{fattah2022smart, van2021new}, including a nested parallel scheme using OpenMP, and integration with the sparse linear solver PARDISO~\cite{gaedke2022parallelized} for the computational core operations.
It does not support distributed-memory parallelism or GPU acceleration. 
%There are various packages that build on top of R-INLA for specific application domains, ... or that have integrated R-INLA in their own framework. 
The INLA\textsubscript{DIST} library \cite{gaedkeIntegrated2024} provides a standalone implementation for SPDE-based spatial and spatio-temporal models~\cite{lindgren2022diffusion}. The C++-based library supports distributed-memory parallelism across the function evaluations, and across the factorization of $\Q_p$ and $\Q_c$ for Gaussian likelihoods, as well as shared-memory parallelism within the matrix operations.
It includes a sequential block-tridiagonal arrowhead (BTA) solver with GPU acceleration for the computational bottleneck operations. A visual overview is provided in Table~\ref{tab:overview_implementation}.

\section{DALIA Framework Innovation}
\label{sec:DALIAframework}
In this section we introduce DALIA and its innovations as a high-performant and scalable spatio-temporal Bayesian modeling framework.
We introduce and define in Table~\ref{tab:notation_daliacs} additional notation and symbols used in this section.

\begin{table}[h]
\scriptsize
\centering
\rowcolors{2}{gray!15}{white} % Alternating row colors start from the second row
\begin{tabular}{p{1.78cm}p{6.25cm}}
\toprule
\textbf{Name} & \textbf{Description} \\
\midrule
% First section: Matrix Parameters
\rowcolor{white} \multicolumn{2}{c}{\textbf{Structured Matrix Parameters}} \\ 
\cmidrule(lr){1-2} % Midrule spanning only the description column
BT/BTA & Block-tridiagonal /with arrowhead structured sparse matrices..\\
$n$ & Number of square blocks in a BTA matrix's main block diagonal, excluding the arrow tip block, equal to $n_t$.\\
$b$ & Size of the square blocks in a BTA matrix's main, upper, and lower block diagonals, excluding the arrowhead blocks, equal to $n_v \cdot n_s$.\\
$a$ & Size of the arrow tip block, equal to $n_v \cdot n_r$.\\
$N$ & Total size of the BTA matrix, equal to $nb + a$.\\
$P$ & Number of time-domain partitions.\\
% Third section: Matrix Types and Algorithms
\cmidrule(lr){1-2}
\rowcolor{white} \multicolumn{2}{c}{\textbf{Computation Symbol and Terminology}} \\ 
\cmidrule(lr){1-2} % Midrule spanning only the description column
$\mathcal{S}_{1},\mathcal{S}_{2},\mathcal{S}_{3}$ & Refer to the different parallelization strategies presented throughout this work. In order; \emph{parallel hyperparameter differentiation}, \emph{parallel precision matrix evaluation}, and \emph{distributed solver}.\\
$G_{S_1},G_{S_2},G_{S_3}$ & Group of processes corresponding respectively to the parallelisation strategies $\mathcal{S}_1$, $\mathcal{S}_2$ and $\mathcal{S}_3$.\\
$\downrsquigarrow^G$ & Symbolize the entrance to a parallel section of the group $G$.\\
$\bigoplus$ & Symbolize an AllReduce operation. \\
\bottomrule
\end{tabular}
\caption{Computational symbols and terms used in this section.}
\label{tab:notation_daliacs}
\end{table}

\subsection{Implementation}
The DALIA framework is developed in Python, leveraging NumPy \cite{harris2020array} and CuPy~\cite{cupy} for array operations, mpi4py~\cite{mpi4py} and NCCL~\cite{nccl_1} for distributed memory communication, SciPy \cite{2020SciPy-NMeth} and cuSPARSE \cite{nvidia_cusparse} for general sparse matrix operations, and Serinv~\cite{serinv_git} as a specialized library for the solution of structured sparse linear system of equations.
These tools were chosen to enable high-performance computing on modern GPU-accelerated systems while maintaining flexibility and scalability.

\subsection{Multivariate Model Formulation}\label{ssec:coreg}
% including permutation scheme 

We derive the precision matrix formulation for the multivariate GP model presented in Sec.~\ref{subsec:MGP} from Eq.~\ref{eq:covariance_LMC}, by separately inverting each matrix component. This can be done efficiently for multivariate processes for any number of response variables $n_v$ as $\vec{\Lambda}$ is lower-triangular and the covariance matrices of the univariate processes are block diagonal. 
Our approach differs from the coregional model formulation used in R-INLA. To simplify the model construction in its user interface, R-INLA relies on artificially introduced linked copies of parameters, inducing an enlarged but sparsified joint precision matrix~\cite{krainski2018advanced}. 
Both formulations yield equivalent results.
For $n_v = 3$ the resulting joint precision matrix $\Q^{n_v}$ of our approach is given by
\begin{equation}
{\small
\Q^{n_v} = 
\left(
\begin{array}{ccc}
   \frac{1}{\sigma_1^2} \Q^1 + \frac{\lambda_1^2}{\sigma_2^2} \Q^2 + \frac{\lambda_3^2}{\sigma_3^2} \Q^3  & -\frac{\lambda_1}{\sigma_2^2} \Q^2 + \frac{\lambda_2 \lambda_3}{\sigma_3^2} \Q^3 & -\frac{\lambda_3}{\sigma_3^2} \Q^3  \\
   -\frac{\lambda_1}{\sigma_2^2} \Q^2 + \frac{\lambda_2 \lambda_3}{\sigma_3^2} \Q^3 & \frac{1}{\sigma_2^2} \Q^2 + \frac{\lambda_2^2}{\sigma_3^2} \Q^3 & - \frac{\lambda_2}{\sigma_3^2} \Q^3 \\
     -\frac{\lambda_3}{\sigma_3^2} \Q^3  & - \frac{\lambda_2}{\sigma_3^2} \Q^3  & \frac{1}{\sigma_3^2} \Q^3
\end{array}
\right)
},
\label{eq:Q_mvn}
\end{equation}
where $\Q^i$ refers to the precision matrix (fixed effects included) of each univariate process as defined in Eq.~\ref{eq:univariate_process}. 
Both the multivariate prior $\Q^{nv}_p$ and the conditional precision matrix $\Q_c^{nv}$ are constructed using Eq.~\ref{eq:Q_mvn}, when respectively relying on $\Q^i_p$ and $\Q^i_c$.
While the joint precision matrices $\Q^{n_v}_{p/c}$ remain sparse, however, and as shown in Fig.~\ref{fig:permutation_coreg}b), the overall BT/BTA  structure is not maintained. 
The dimensions of the joint matrices are the sum of the dimensions of the univariate precision matrices, i.e. dim$(\Q^{n_v}) = N = n_v (n_s n_t + n_r)$, where $n_v$ is the number of univariate processes, $n_s$ the spatial mesh size, $n_t$ the number of time steps and $n_r$ the number of fixed effects for each univariate process. For simplicity, we assume them to be the same. If their dimensions differed one would additionally require projection mappings between the different meshes for their joint matrix assembly.
Each of the $\Q^i_{p}$ consist of the sum of sparse Kronecker products arising from the spatial and temporal mesh discretization~\cite{lindgren2022diffusion} and has a separate set of unknown hyperparameters which was omitted in the notation above for readability. 
%They need to be recomputed for every new hyperparameter configuration of $\th$ and are joint with the prior precision matrix of the fixed effects.

% \subsubsection{Permutation Strategy}\label{sec:permute_model}
\subsubsection{Permutation}\label{sec:permute_model}

By construction, $\Q^{n_v}$ is ordered by response variable, which are internally ordered by time step.
To recover the BT/BTA sparsity patterns, as in the univariate model, we reorder $\Q^{n_v}$ after construction such that all parameters of all response variables that are associated with the first time step are aggregated in the top left corner of the matrix. Concatenating the $n_v$ spatial domains results, for the first time step, in an enlarged diagonal blocksize $b = n_v n_s$. 
% This process is repared for the second time step, permuting all associated variables to a second diagonal block which is internally ordered by response variable. 
This procedure is repeated for all subsequent time steps, recovering a block tridiagonal structure.
We additionally accumulate all fixed effects at the end,
giving rise to the coregional model BT and BTA  sparsity pattern Fig.~\ref{fig:permutation_coreg}c). 
% with enlarged diagonal block sizes but the same number of time steps as in the univariate models, see Fig.~\ref{fig:permutation_coreg}c). 
As each of the univariate processes has their unique set of hyperparameters, we cannot directly construct the permuted matrix but require an efficient permutation strategy to be applied during every $f_{\text{obj}}$ evaluation. 
This is realized by computing and storing the necessary permutation of the sparse matrix once, which is then directly applied to the matrix data array without the need to modify or recompute the row and column indices.
%\textit{transition clearly missing. lets write it once the time domain partitioning is there.}

\begin{figure}[t]
    \centering
    \includegraphics[width=\columnwidth]{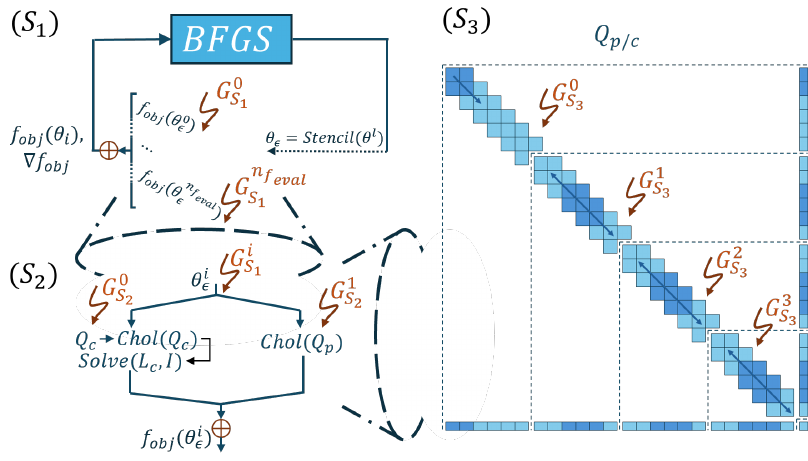}
    \caption{Workflow diagram of the three nested parallelization strategies:
    $\mathcal{S}_1)$ Parallel evaluations of the objective function via a central difference scheme for the differentiation of the hyperparameters;
    $\mathcal{S}_2)$ Parallel construction of prior and conditional precision matrices, Cholesky factorization, and linear system solving;
    $\mathcal{S}_3)$ Parallel linear system solving, including parallel Cholesky factorization of the BTA matrix and parallel triangular solve, distributed across 4 processes. 
    The $\downrsquigarrow^G$ symbol denotes entry into an MPI parallel section, with superscripts indicating the group/process entering the section.}\label{fig:parallelisation_strategies}
\end{figure}

\subsection{Time-domain Partitioning}\label{sec:time_domain_part}
The reordering of the temporal and spatial domains, leading to the recovery of the BT/BTA structured sparsity pattern, allows us to leverage GPU-accelerated dense operations.
This further enables the usage of structured linear solvers at the block level with a computational complexity of $O(n \cdot b^3)$.
% This separation gives rise to a BT (resp. BTA) structured sparsity pattern, which, from a computational standpoint, allows us to leverage GPU-accelerated dense operations at the block level with a computational complexity of $O(n*b^3)$.
This approach overcomes the traditional challenges of general sparse solvers by breaking sequential dependencies and exposing the large amount of parallelism required to perform efficient computations on hardware accelerators.
However, it also comes with a trade-off in the representation of the underlying data.
Since it densifies the blocks that lie inside the sparsity pattern, the memory footprint increases from $O(nnz)$ (for general sparse matrix representation) to $O(n \cdot b^2)$.
This increase can limit the representable model sizes, as each model has to fit on a single accelerator in its densified BT/BTA form.
To overcome this issue, we introduce a time-domain partitioning by distributing slices of the structured matrix to different processes.
We then utilize the distributed routines of the structured sparse solver library \textit{Serinv} to operate in a distributed manner on these partitions.

\subsection{Nested Distribution Layers}
We introduce the triple-layer, nested, parallelization strategies, denoted by $\mathcal{S}_1$, $\mathcal{S}_2$ and $\mathcal{S}_3$, and ordered from outer to inner layers.
The first two layers of parallelism leverage specific characteristics of the methodology, while the third and innermost layer relies on the time-domain partitioning of the precision matrices.

\subsubsection{Parallel objective function evaluations}
The outermost parallelization strategy $\mathcal{S}_1$ consists of the parallel function evaluations of $f_{\text{obj}}$ to approximate the gradient as described in Eq.~\ref{eq:FD_scheme} and is illustrated in Fig.~\ref{fig:parallelisation_strategies}a). 
Using a first-order central difference scheme, while also evaluating $f_{\text{obj}}$ in its central point, results in a total of  $n_{f_{eval}} = 2 \cdot \text{dim}(\th) + 1$ parallelizable function evaluations. 
%This approach allows us to distribute the computationally expensive tasks of computing the objective function values of $f_{obj}$ at each evaluation point of the central difference scheme.  
The trivariate coregional model presented in Sec.~\ref{ssec:coreg} includes 15 hyperparameters, for a total of $n_{f_{eval}^{coreg}} = 31$ parallel function evaluations in every BFGS iteration.
The computed function values are aggregated using the AllReduce operation, represented by the $\bigoplus$ symbol in Fig.~\ref{fig:parallelisation_strategies}a). 
When the number of processes is a divisor of $n_{f_{eval}}$, this strategy achieves ideal theoretical scaling as the function evaluations are embarrassingly parallel. 
The only additional overhead is the computation of the central difference matrix and the reduction operation at the end of all function evaluations. 
The experimental results presented Sec.~\ref{sec:comparison_soa}, confirm this behavior.
The first parallelization layer is saturated once the number of processes exceeds $n_{f_{eval}}$.

\subsubsection{Parallel precision matrix decompositions}\label{sssec:pqeval}
We consider the 4 components of $f_{\text{obj}}(\th^l)$ individually.
The first two are computationally negligible, while the last two are computationally intensive.
% We separate the evaluation of the subcomponents within $f_{\text{obj}}(\th)$ in parts. The first three and the last term, as the first two terms are negligible in terms of comptuational cost. 
The evaluation of the computationally intensive parts within each objective function evaluation can be parallelized, leading to the second layer of parallelism $\mathcal{S}_2$.
Therefore, $\mathcal{S}_2$ parallelizes the construction and factorization of $\Q_p$ and $\Q_c$, and the solution of the linear system associated with the latter.
% Due to the difference in sparsity pattern of $\Q_p$ and $Q_c$, a slight load imbalance is introduced at this level.
Due to the difference in sparsity pattern between $\Q_p$ and $\Q_c$, and the additional triangular solve operation on the conditional precision matrix, a slight load imbalance is introduced at this level.
The respective computational cost of the factorization of a BT (BTA) matrix is $O(n \cdot b^3)$ ($O(n \cdot (b^3+a^3))$), while the asymptotic load balancing ratio is given by $r_Q = \frac{a^{3}}{b^{3}}$.
This second parallelization layer is illustrated in Fig.~\ref{fig:parallelisation_strategies} and nested within each parallel function evaluation, i.e., $\mathcal{S}_1$, resulting in a total of $n_P = 2 \cdot n_{f_{eval}}$ parallelizable sections.

\subsubsection{Parallelization through the temporal domain}
% two main points
% parallelism within factorization 
% leverage particular sparsity pattern
As the dimensions of the modeled spatio-temporal domains increase, our framework introduces a novel third layer of distributed memory parallelism: $\mathcal{S}_3$.
This layer builds on top of the time-domain partitioning introduced in Sec.~\ref{sec:time_domain_part} and uses a distributed-memory, structured-sparse, linear solver from the \textit{Serinv} library, allowing DALIA to scale to larger models.
% but located at the computational core of the method, representing about $90\%$ of the runtime for relevant workloads, also allow to push further the library' performances.
The distributed solver relies on a nested dissection scheme to break the sequential dependencies inherent to the factorization and selected inversion routines, which can induce a computational load imbalance depending on the temporal partitioning.
% The distributed solver can induce a computational load imbalance depending on the temporal partition.
% In practice, and to break the sequential dependencies inherent to the factorization and selected inversion routines, the library relies on a nested dissection scheme.
The nested dissection scheme introduces an added workload for the middle partitions of the temporal domain, leading to the computational imbalance.
This effect can be mitigated in the time-domain partitioning strategy by assigning more time steps to the partitions at the boundaries of the temporal domain.
Load balancing is further discussed in Sec.~\ref{sssec:micro_solver}.

\subsection{Distributed Triangular Solve}\label{ssec:dist_solve}
While the \textit{Serinv} library includes distributed-memory implementations for matrix factorization and selected inversion, it lacks a triangular solver for BT and BTA matrices.
Therefore, we developed our own solution, the routine $PPOBTAS$, leveraging the same nested dissection scheme.

\begin{table}[t]
% \scriptsize
\centering
\resizebox{\columnwidth}{!}{%
% \rowcolors{2}{gray!15}{white} 
\begin{tabular}{ccccc}
\toprule
 % & \textbf{dim($\theta$)$/n_v$}  & \textbf{$n_s/n_r$} &  \textbf{$n_t$} & \textbf{$N$} \\
 & \textbf{dim(\bm{$\theta$})$\bm{/n_v}$} & \textbf{$\bm{n_s/n_r}$} & \textbf{$\bm{n_t}$} & \textbf{$\bm{N}$} \\
\midrule
$MB_1$ & 4/1    &   4002/6     &    250  &   1\,000\,506    \\
$MB_2$ & -/1   &    1675/6    &  128 - 2048   &  214\,406 - 3\,3430\,406  \\
% \hline
\cmidrule(rl){1-5}
% \cmidrule(lr){1-2}
$WA_1$  & 15/3     &   1247/1    &   2 - 512    &   7\,485 - 1\,915\,395   \\
\multirow{2}{*}{$WA_2$} & \multirow{2}{*}{15/3}   &    [72, 282,    &    \multirow{2}{*}{48}    &   \multirow{2}{*}{10\,371 - 645\,843}    \\
& & 119, 4485]/1\\
% \hline
% \cmidrule
\cmidrule(rl){1-5}
$SA_1$ &  15/3    &   1675/1      &   192    &   964\,803    \\
\cmidrule(rl){1-5}
$AP_1$ &  15/3    &   4210/2      &   48    &   606\,246    \\
\bottomrule
\end{tabular}
} % end resizebox
\caption{Datasets used in the performance evaluations of the DALIA framework. Here, dim($\th$) describes the number of hyperparameters in the model, $n_v$ the number of univariate processes, $n_s$ the spatial mesh size per process, $n_r$ the number of fixed effects, $n_t$ the number of time steps and $N$ the total matrix dimension.}
\label{tab:datasets}
\end{table}

\subsection{Sparse to Structured Dense Matrix Mapping}
For every new configuration of $\th$, updated precision matrices need to be computed. Their assembly is based on a sum of sparse Kronecker products and gives rise to positive-definite sparse matrices.
To enable the distributed solver to efficiently operate on the underlying BT/BTA block structure of these matrices, the non-zero entries in the precision matrices need to be mapped efficiently to the solver's workspace.
To overcome native Python language limitations when it comes to efficiently operating on sparse matrices, we implemented custom CUDA kernels, effectively bringing the complexity of this memory operation from $O(n \cdot b^2)$ to $O(nnz)$.
The mapping complexity is further reduced to $O(\frac{nnz}{P})$ where $P$ is the number of processes in the group $G_{S_3}$.

\section{Performance Evaluation}
\label{sec:results}

\subsection{Evaluation Methodology and Hardware}\label{sec:eval_hw}

In Table~\ref{tab:datasets} we present an overview of the dimensions of the different models and datasets used for our performance analyses.
We conducted our GPU experiments at the Swiss national computing center (CSCS) on the Alps supercomputer equipped with NVIDIA GH200 superchips (Grace CPU and Hopper GPU), using the NVIDIA Collective Communications Library (NCCL) at the solver-level (mostly intranode communication) and MPI otherwise.
We run the R-INLA package on the Intel Xeon Platinum 8470 ``Sapphire Rapids'' CPU (2TB partition) of the Fritz supercomputer at the Erlangen National High Performance Computing Center (NHR@FAU).
We used the sparse direct solver PARDISO backend for R-INLA and empirically determined the most favorable nested OpenMP thread configuration for R-INLA, following the guidelines provided in~\cite{gaedke2022parallelized}. We chose  $\mathcal{S}_1=9$, $PARDISO_{\text{omp}}=8$ for case study $MB_1$ and $\mathcal{S}_1=8$, $PARDISO_{\text{omp}}=8$ for case study $WA_{1,2}$ and $SA_1$. 
Here $\mathcal{S}_1$ represents the number of groups performing the finite difference function evaluation in parallel, and $PARDISO_{\text{omp}}$ the number of threads per sparse solver instance within each group $\mathcal{S}_1$.

In our benchmarks, we stop the comparison with R-INLA once it becomes more than two orders of magnitude slower than our approach. Since R-INLA is restricted to shared memory parallelism we always run it in its most performant configuration.
Unless stated otherwise all timings reported are \emph{per iteration} of the minimization procedure introduced in Sec.~\ref{sec:inla}.

\subsection{Comparison to State-of-the-art}\label{sec:comparison_soa}
% \subsubsection{Comparison to state-of-the-art}\label{sssec:eval_ps1s2}

\begin{figure}[t]
  \centering
  \includegraphics[width=0.9\columnwidth]{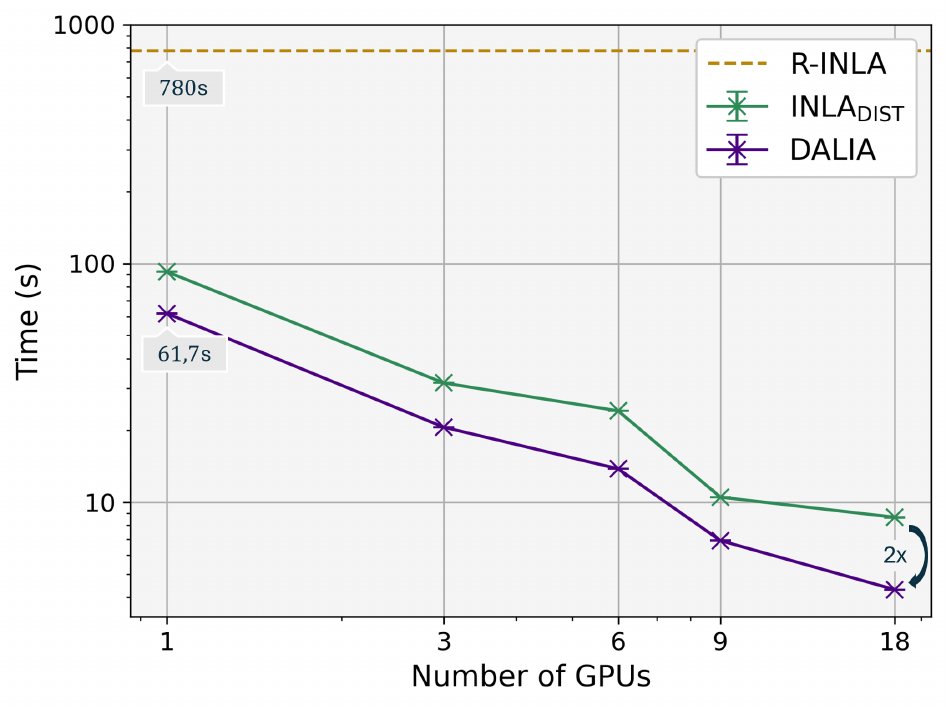}
  \caption{Strong scaling comparison of our proposed framework DALIA to the reference INLA\textsubscript{DIST} and R-INLA libraries. Scaling is shown from $1$ to $18$ GPUs, showcasing the parallelisation through the function evaluations and precision matrix decompositions. The results are for a univariate spatio-temporal model made of $250$ time-steps and a spatial mesh of $4002$ nodes.}
  \label{fig:peval_microbench}
\end{figure}

For our first benchmark we use a large-scale, univariate, spatio-temporal model (dataset: $MB_1$) from~\cite{gaedkeIntegrated2024} and showcase the parallelization through strategies $\mathcal{S}_1$ and $\mathcal{S}_2$ as INLA\textsubscript{DIST} does not support multivariate models or solver-level distributed memory parallelism. 
We present a strong scaling comparison of DALIA and INLA\textsubscript{DIST} in Fig.~\ref{fig:peval_microbench}, and include a runtime comparison to the R-INLA package.
On a single GPU, DALIA (resp. INLA\textsubscript{DIST}) offers a $12.6$-fold (resp. $8.4$-fold) speedup over the R-INLA package.
When scaling up to $18$ GPUs, DALIA offers a twofold speedup over INLA\textsubscript{DIST}, achieving a parallel efficiency $\eta=79.7\%$ (INLA\textsubscript{DIST} achieving $\eta=59.3\%$).
On $18$ GPU, DALIA offers a $180$-fold speedup over the parallel, shared-memory, R-INLA package, bringing the runtime required per iteration from $780s$ to $4.3s$.

\begin{figure}[t]
  \centering
  \includegraphics[width=0.9\columnwidth]{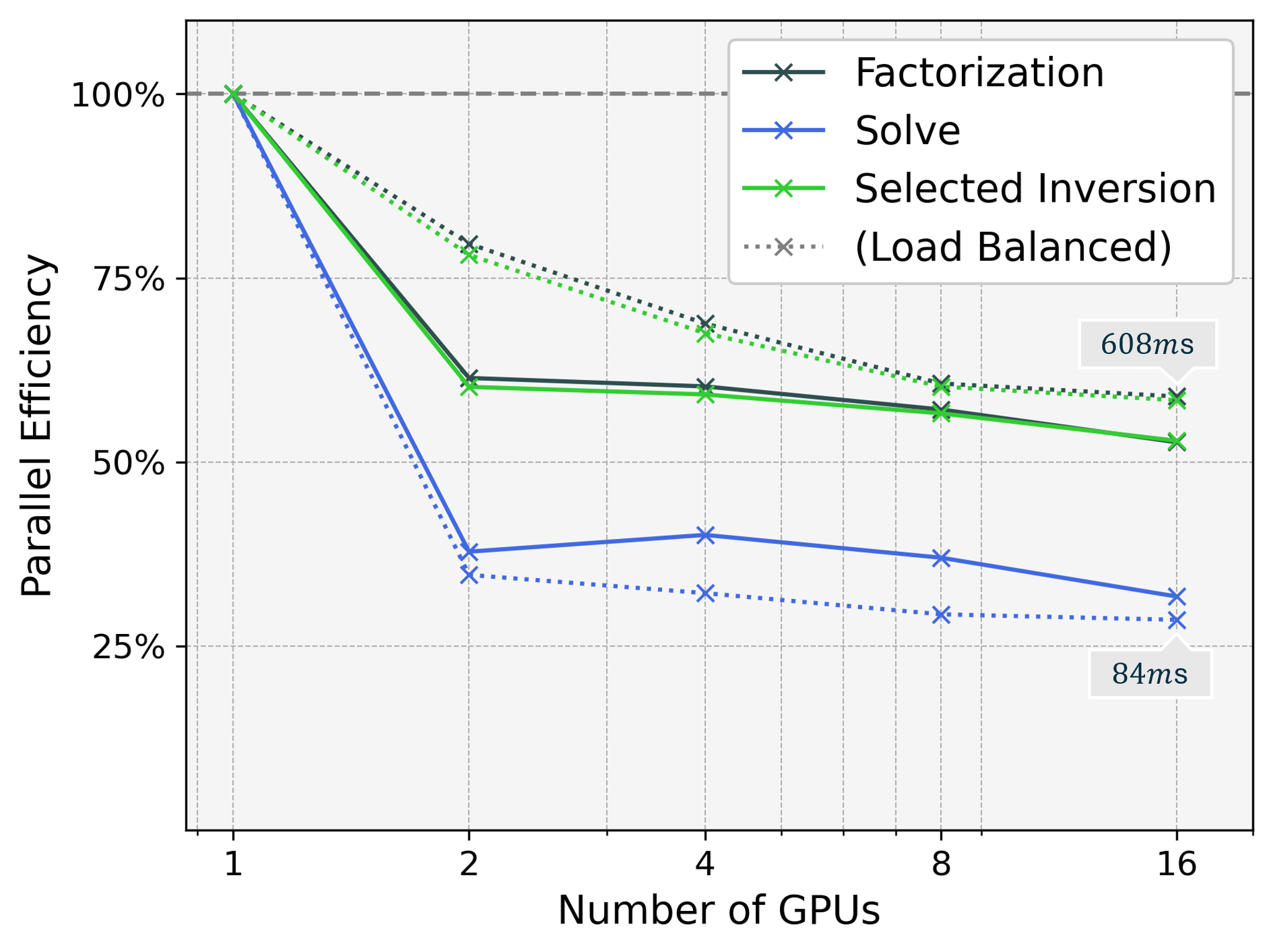}
  \caption{Weak scaling parallel efficiency of the distributed memory Cholesky factorization, selected inversion and triangular solve routines. Results are shown for a univariate spatio-temporal model made of $1675$ spatial nodes, each process solving for $128$ time-steps ($MB_2$).}
  \label{fig:solve_microbench}
\end{figure}

% \subsubsection{Solver microbenchmarks]}\label{sssec:micro_solver}
\subsection{Solver Microbenchmarks}\label{sssec:micro_solver}

We perform a weak scaling analysis through the time domain (dataset: $MB_2$) of the \textit{Serinv} distributed solver routines we have integrated in our framework as well as the distributed triangular solve routine we developed. 
We use the NCCL communication library and run the solver with and without load-balancing.
We present our results in Fig.~\ref{fig:solve_microbench}.
The distributed factorization and selected inversion routines respectively achieve $52.6\%$ and $52.8\%$ parallel efficiency when scaling from $1$ to $16$ GPUs.
When using an appropriate load balancing factor (here $lb=1.6$) for the first partition, the parallel efficiency of these routines improves to $58.8\%$ and $58.3\%$, respectively .
The load balancing plays, however, the most important role when scaling from $1$ to $2$ processes, as it is shown to reduce the runtime of both factorization and selected inversion routines by about $30\%$ in these cases.
Finally, the newly implemented triangular solve achieves a parallel efficiency of $31.6\%$ on 16 GPUs.
% Contrary to other two routines it performs worse under load balancing, as the work for this routine is the same for all partitions. 
Contrary to the other two routines it performs worse under load balancing.
In practice, the triangular solve routine is about an order of magnitude faster than both factorization and selected inversion. Therefore, the improvement in runtime by an adapted load-balancing ratio in the latter two completely compensates for the performance loss during the triangular solve.

Even though our microbenchmarks clearly show an advantage in using load balancing this is not always possible in practice due to memory constraints.
In the following large scale benchmarks, we will use load balancing when possible and fall-back to an even distribution when the device memory does not allow for storing a larger partition for the first process.

\begin{figure*}[t]
    \centering
    \includegraphics[width=\textwidth]{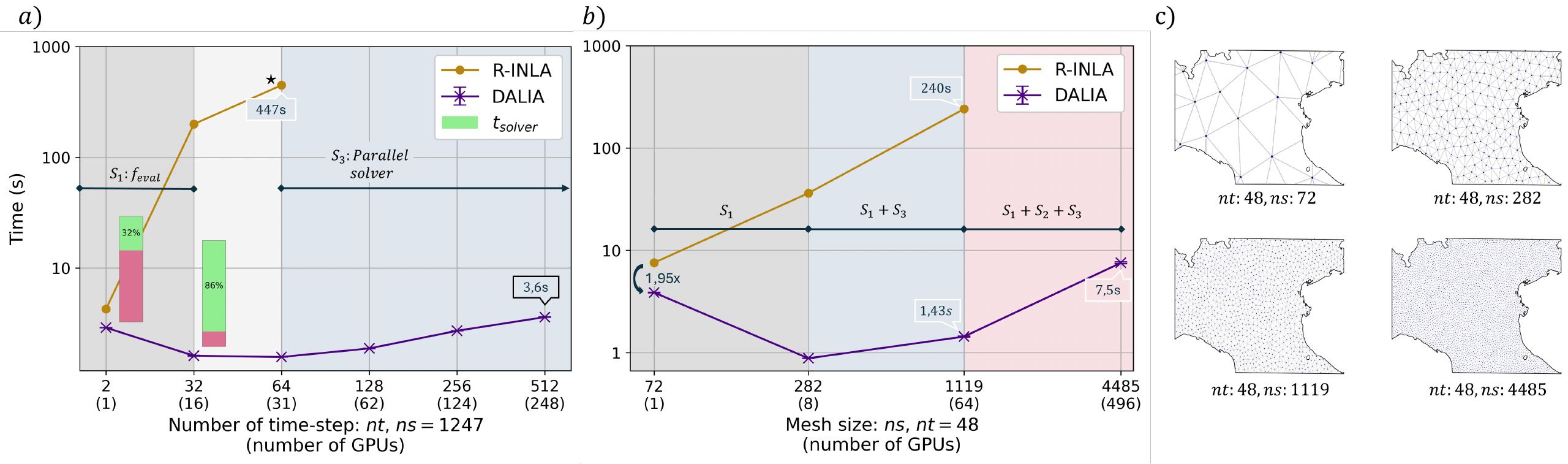}
    \caption{Weak scaling results of the DALIA framework on the Grace-Hopper GPUs of the CSCS-Alps supercomputer, references runtimes for the R-INLA package achieved on the Fritz supercomputer of the FAU computing center. $a)$ Weak scaling through temporal discretization of a trivariate coregional model, from $2$ time-steps ($1$ GPU) to $512$ time-steps ($248$ GPUs). $b)$ Weak scaling through spatial mesh refinement, from $72$ mesh nodes ($1$ GPU) to $4485$ nodes ($496$ GPUs). $c)$ Spatial meshes at different refinement levels on northern region of Italy. }
    \label{fig:weak_scaling}
\end{figure*}

\subsection{Weak Scaling at the Application Level}~\label{sec:weakapp}
In this section, we apply our framework to trivariate Gaussian process models. 
We first present a scaling analysis through the time domain, Section~\ref{sec:ws_time}, and then a spatial scaling analysis using mesh refinement, Section~\ref{sec:ws_space}.
In both analyses, we scale a matrix parameter (the number of diagonal blocks and the diagonal block size, respectively) while simultaneously increasing the available computing resources. 
We always distribute the added computational resources across the most efficient unsaturated parallelization strategy.
%However, we perform a "smart" distribution of these computational resources by allocating them to the most efficient parallelization strategy.
This results in favoring parallelizing $\mathcal{S}_1$ over $\mathcal{S}_2$ and $\mathcal{S}_3$.
We only distribute through $\mathcal{S}_3$ first when memory constraints force us to distribute the block dense precision matrices. 
%We expect our framework to scale linearly in time, thus doubling the number of processes, when doubling the number of time steps and cubically in space, that  when doubling the number of spatial mesh nodes.
% however,
We observe that for small problem sizes, the majority of the runtime is not spent in the solver but mainly in the precision matrix construction whose runtime does not exhibit the same scaling behavior but becomes negligible for larger models. 
Our application can therefore exhibit superlinear scaling for small problems.
%, as the workload per processes go down in the embarassingly parallel strategy $\mathcal{S}_1$.
This is explained further in the following sections.

\subsubsection{Scaling in time}\label{sec:ws_time}
We perform a weak scaling analysis in the time domain using dataset $WA_1$. 
The results are presented in Fig.~\ref{fig:weak_scaling}a). With two time-steps ($1$ GPU), our framework demonstrates a $1.48$-fold speedup over R-INLA. From $32$ time-steps ($16$ GPUs) onward, our framework outperforms R-INLA by more than two orders of magnitude. 
%We don't conduct further benchmarks on R-INLA based on constraints introduced in section~\ref{sec:eval_hw}
We further scale our implementation up to $512$ time-steps ($248$ GPUs), achieving a 124-fold speedup over R-INLA while operating on a model eight times larger (comparison to point "$\star$").
As we scale our application from 1 to 16 GPUs through $\mathcal{S}_1$, we observe a superlinear scaling trend. As introduced in Sec.~\ref{sec:weakapp}, this behavior is due to a reduction of the workload per process as $\mathcal{S}_1$ is embarrassingly parallel and does not depend on the scaling parameter $n_t$. 
From 64 time-steps onward, as the solver routines become the major contributor to the runtime (approximately $90\%$), we observe the expected correlation between time discretization and application scaling. We provide a stacked bar plot next to the initial data points, illustrating the percentage of total time spent in the solver (green portion of the bar). 

\subsubsection{Scaling in space}\label{sec:ws_space}
We perform a weak scaling analysis in the spatial domain using dataset $WA_2$. 
The results are presented in Fig.~\ref{fig:weak_scaling}b) and the 4 levels of mesh refinement used in Fig.~\ref{fig:weak_scaling}c). Using the coarsest spatial mesh, DALIA shows a 1.95 times speedup over R-INLA. We then proceed to scale the application to 8 GPUs through $\mathcal{S}_1$. Once again, and as explained in the last section, we observe a superlinear trend as the workload per process decreases. From 282 mesh nodes to 1119 mesh nodes and as the model, where the model does not fit on a single device anymore at the solver level, we then parallelize through $\mathcal{S}_3$, leading to a combined strategy $\mathcal{S}_1+\mathcal{S}_3$ and a 168-folds speedup over R-INLA.
Finally, we scale up to 496 GPUs by combining all three parallelization strategies $\mathcal{S}_1+\mathcal{S}_2+\mathcal{S}_3$, achieving a parallel efficiency of $\eta = 51.2\%$.

\subsection{Strong Scaling at the Application Level}~\label{sec:strongapp}
\begin{figure}[t]
  \centering
  \includegraphics[width=0.9\columnwidth]{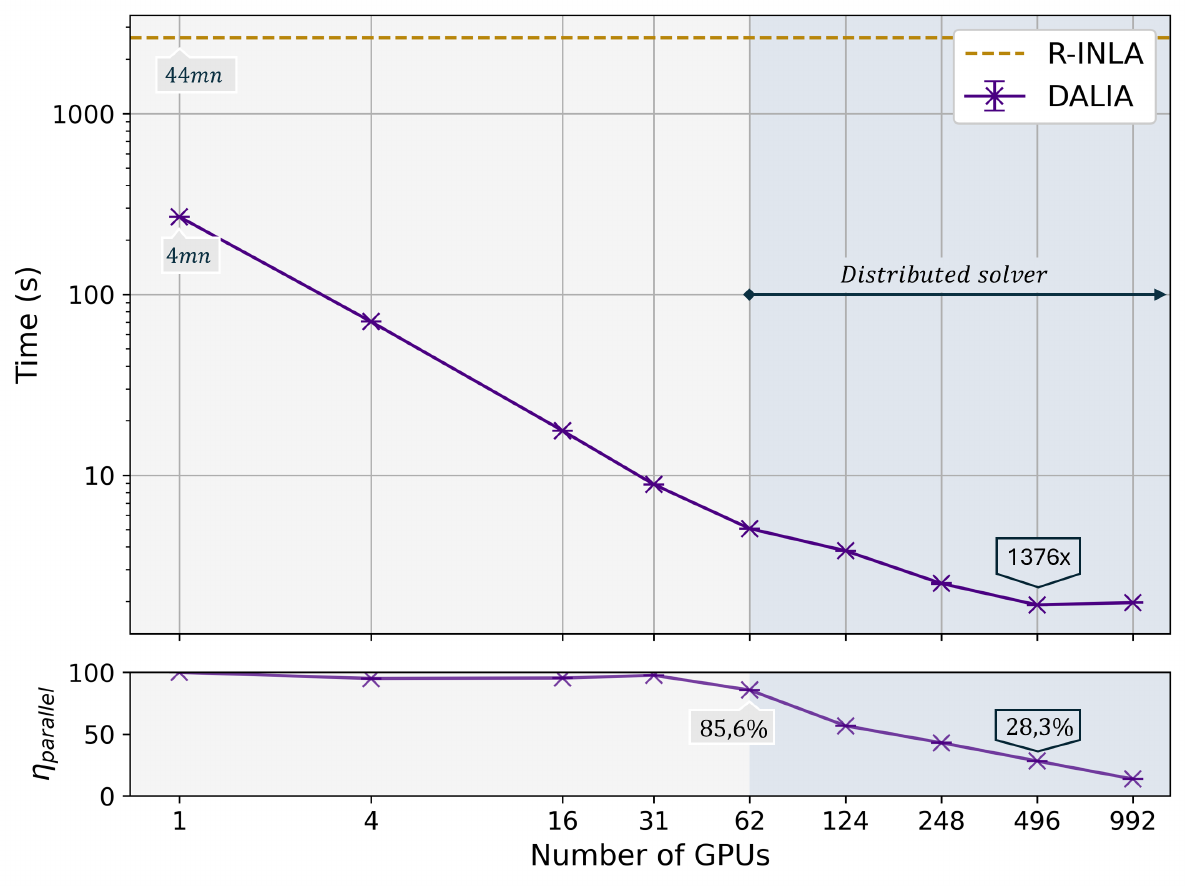}
  \caption{Strong scaling evaluation of the DALIA framework on a trivariate coregional model comprising $1675$ spatial mesh nodes per univariate process and modeled over $192$ days. Runtime comparison to the R-INLA package (top-figure) and achieved parallel efficiency (bottom-figure).}
  \label{fig:strong_scaling}
\end{figure}

Certain applications, such as data disaggregation for daily forecasting, are computationally time-critical due to fixed and recurrent deadlines for producing scientific results. 
%In such cases, our framework enables these applications to harness modern distributed-memory and GPU-accelerated computing environments, thereby reducing inference times. 
Here, we provide a strong scaling evaluation of our framework on the trivariate spatio-temporal models described in dataset $SA_1$. The results are presented in Fig.~\ref{fig:strong_scaling}. On a single GPU, our application takes about 4mn per iteration while R-INLA takes more than 40mn. Our application shows near-perfect scaling efficiency up to 31 GPUs and reaches a parallel efficiency of $\eta = 85,6\%$ on 62 GPUs.
When parallelizing through the last strategy $\mathcal{S}_3$ to scale further, our application exhibits peak performances at 496 GPUs, with a parallel efficiency of $28.3\%$ and a three orders of magnitude speedup over R-INLA.

%\section{Scientific results}
\section{Application to Air Pollution Dataset}
\label{sec:scientific_results}

\begin{figure}[t]
  \centering
  \includegraphics[width=\columnwidth]{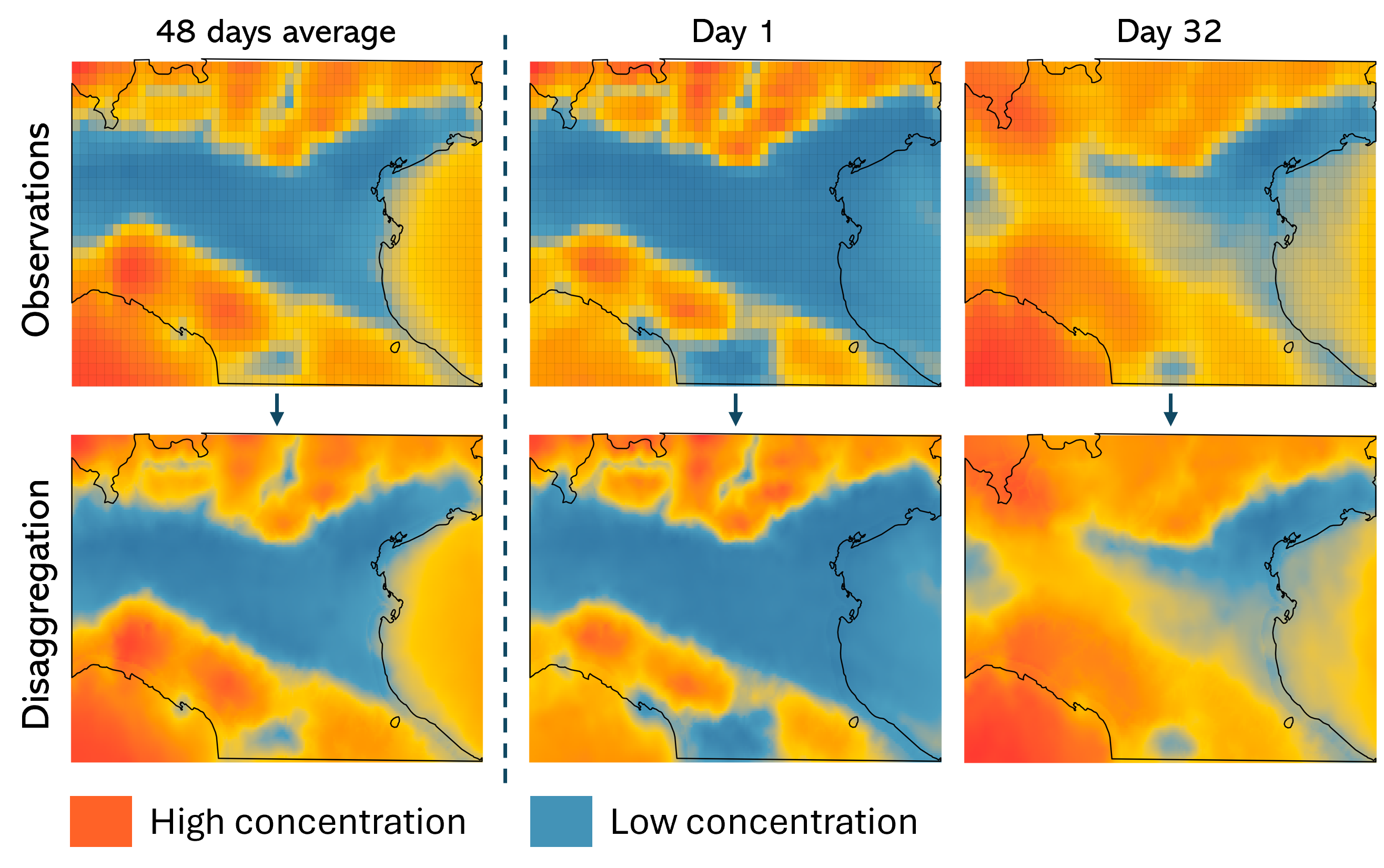}
  \caption{Spatial downscaling results using the proposed multivariate spatio-temporal LMC. The top row shows the original coarse-resolution input for Ozone concentrations. The bottom row presents the corresponding high-resolution predictions.}
  \label{fig:scientific_results}
\end{figure}

We apply our framework to a multivariate spatio-temporal coregionalization model on air pollution, jointly modeling PM$_{2.5}$, PM$_{10}$, and O$_3$\cite{YaweiEtAl2023, rodriguez2025multivariatedisaggregationmodelingair}. 
We used hourly pollutant measurements provided by the Copernicus Atmosphere Monitoring Service (CAMS) \cite{copernicus_data} with a spatial resolution of $0.1^\circ$ ($\approx 10 \text{km}$). 
We aggregated the data to daily resolution for analysis, covering a $122\, 350\,\text{km}^2$ region in northern Italy with 4210 spatial locations and 48 days of observations, starting from January 1st, 2022, see also $AP_1$ in Table~\ref{tab:datasets}.
Using our framework we perform spatial downscaling to a resolution of $0.02^\circ$ ($\approx 2 \text{km}$), increasing spatial detail 25-fold.
%This is achieved through a spatio-temporal LMC incorporating altitude as an additional covariate, following the structure described in Equation~\ref{eq:multivariate_model}. 
The results of the downscaling process of the multivariate GP for ozone  (O$_3$) concentration are presented in Fig.~\ref{fig:scientific_results}. 
% The first row shows the input data, while the second row displays the disaggregated output. 
The first row shows the coarse satellite measurements, while the second row displays the disaggregated estimated mean. 
The first column represents the averaged data over the entire time period and reveals consistently high average ozone levels in oceanic areas. 
% Nevertheless, the model effectively captures temporal anomalies, including unusually high concentrations in the western region on January 1 (day 1) and low concentrations in the eastern region on February 1 (day 32).
% The temporal-domain discretization reveals that averaged data over the entire time period fails to represent the behavior of individual days, including unusually high concentrations in the western region on January 1 (day 1) and low concentrations in the eastern region on February 1 (day 32).
A time-resolved analysis reveals that the averaged data over the entire time period fails to represent the behavior of individual days, including unusually low concentrations in the eastern region on January 1 (day 1) and high concentrations in the western region on February 1 (day 32).
This behavior is, however, accurately captured by our spatio-temporal approach, underpinning the need for efficient implementations that allow for spatio-temporal instead of only spatial multivariate modeling. 
Our approach also provides posterior estimates for the models' interpretable parameters, such as the effect of elevation on pollutant concentrations. Specifically, a 1 km increase in elevation is associated with reductions of $0.45~\mu g/m^3$ in PM$_{2.5}$, with the 0.025 and 0.975 quantiles estimated at $-0.51$ and $-0.40$, respectively; a reduction of $0.55~\mu g/m^3$ in PM$_{10}$, with corresponding quantiles at $-0.62$ and $-0.49$; and an increase of $1.27~\mu g/m^3$ in O$_3$, with quantiles of $1.20$ and $1.34$. 
% In addition, the model provides linear correlations between pollutants, shown in Table~\ref{tab:corr_matrix}, revealing a strong positive correlation ($0.97$) between PM$_{2.5}$ and PM$_{10}$, and moderate negative correlations with O$_3$: $-0.61$ for PM$_{2.5}$ and $-0.63$ for PM$_{10}$.
In addition, we obtain the following linear correlations between pollutants, revealing a strong positive correlation ($0.97$) between PM$_{2.5}$ and PM$_{10}$, and moderate negative correlations with O$_3$: $-0.61$ for PM$_{2.5}$ and $-0.63$ for PM$_{10}$.

\section{Conclusion and Outlook}
\label{sec:conclusion}
In this work, we present DALIA, a highly scalable implementation of the INLA methodology for multivariate  spatio-temporal Bayesian modeling. 
Our contributions include a computationally advantageous expression of coregional models, a hierarchical three-layer distributed memory parallelization scheme, a time-domain decomposition of the spatio-temporal model formulation, the development of a distributed-memory triangular solve routine based on the \textit{Serinv} library permutation scheme, and the integration of the latter in our framework. 
%We evaluated our framework DALIA on univariate spatio-temporal models against the reference implementations $INLA_{\text{DIST}}$ and R-INLA , achieving a speedup of 2x (resp. 181x) on 18 GPUs.
% We then demonstrated our efficent formulation of multivariate, coregional, spatio-temporal model and performed weak scaling analysis through both spatial and temporal domain.
% We compared the performances of our application to the only other capable package R-INLA.
% When performing weak scaling in the temporal domain we show a 124-fold speedup while operating on a 8-times larger temporal domain.
% Our weak scaling in spatial-domain refinement reveals a 168-fold speedup when running on 64 GPUs and achieve a parallel efficiency of $\eta = 51.2\%$ on 496 Hopper GPUs.
% Finally, we conducted strong scaling experiments and reached a peak three order of magnitude speedup over the R-INLA package when scaling to 496 Hopper GPUs.
We demonstrated the scalability of our new framework in weak scaling analyses, through both the temporal as well as the spatial domain. In the former we exhibit a 124-fold speedup while operating on a 8-times larger temporal domain over the reference library, R-INLA. 
Our weak scaling experiments in the spatial domain reveals a 168-fold speedup over the reference when running on 64 GPUs and achieves a parallel efficiency of $\eta = 51.2\%$ on 496 Hopper GPUs.
We conducted strong scaling experiments reaching a speedup of
three orders of magnitude over the R-INLA package when scaling to 496 Hopper GPUs.
% These innovations enable DALIA to surpass the reference INLA\textsubscript{DIST} framework on large scale spatio-temporal models and to scale the INLA methodology on multivariates spatio-temporal models by 1 order of magnitude in the tackled problem sizes while achieving a 2 orders of magnitudes speedup over the state-of-the-art R-INLA package.
% Additionally, we conducted weak and strong scaling analyses on both real and synthetic datasets, surpassing the state-of-the-art by a factor of YY and achieving a sustained performance of ZZ.
We applied our new framework to model air pollution in northern Italy over 48 consecutive days, demonstrating its ability to provide fast and realistic inference results on real-world data.
%, jointly modeling PM${2.5}$, PM${10}$, and O$_3$ concentrations. Using daily aggregated CAMS data at 0.1$^\circ$ resolution, we downscale to 0.02$^\circ$, enhancing spatial detail 25-fold. As shown in Fig.~\ref{fig:strong_scaling}, the model captures both large-scale patterns and day-specific anomalies in O$_3$ levels that are missed by time-averaged data. 
%It also provides interpretable insights, such as the effect of elevation and pollutant correlations. Although demonstrated here for air quality, the same approach extends naturally to spatio-temporal applications in fields like epidemiology and ecology.
% In this paper, we illustrate how to use DALIA for air pollution modeling. This methodology can also be used to model spatio-temporal processes in a wide range of scientific disciplines such as disease mapping or species distribution modeling

%Although 
As DALIA pushes the boundaries of the INLA methodology in terms of computational scale, it also reveals new challenges. For instance, the lack of parallelization at the spatial domain level limits the feasible dimension of the spatial mesh on current hardware devices.
This limited our application study to a large region instead of covering entire countries. 
% Specifically, our distributed solver requires at least three block-rows to fit on a single node. 
% We demonstrated this with a spatial field on the northern region of Italy, where diagonal blocks of (TODO verify/change) 15GB allowed us to fit approximately 8 blocks per H100 GPU. 
%We demonstrated current capabilities with a temporal discretization over 48 days and spatial field on the northern region of Italy with a resulting blocksize of 13455.
%This allowed us to fit 6 of these blocks in the 96HMB2 memory of the H100 GPUs.
%However, for larger spatial fields, such as those covering an entire country like Portugal or Italy, diagonal blocks can exceed 100GB, approaching the limits of current hardware. 
This highlights the need for future work on the parallel computation on spatial domains, to further extend the scalability of our framework.

\section*{Acknowledgments}
This work was supported by the Swiss National Science Foundation (SNSF) under grant $\mathrm{n^\circ}$ 209358 (QuaTrEx), grant $\mathrm{n^\circ}$ 200021 (NumESC), and by the Platform for Advanced Scientific Computing in Switzerland (BoostQT). We acknowledge the scientific support and HPC resources from CSCS under projects sm96 and lp16 as well as the Erlangen National High Performance Computing Center (NHR@FAU) of the Friedrich-Alexander-Universität Erlangen-Nürnberg (FAU) under the NHR project 80227.

\bibliographystyle{IEEEtran}
\bibliography{sample-base.bib}

\end{document}